\pdfoutput=1

\documentclass[aps,pra,10pt,twocolumn,showpacs,superscriptaddress,times,floatfix]{revtex4-1}

\usepackage{amsmath}
\usepackage{latexsym}
\usepackage{amssymb}
\usepackage{bm}
\usepackage{graphics,epstopdf,subfigure,graphicx}
\usepackage{color}

\newcommand{\braket}[1]{\langle #1 \rangle}

\newcommand{\tr}{\mbox{tr}}

\input{Qcircuit}

\begin{document}

\author{Hugo Cable}
\affiliation{Centre for Quantum Photonics, H. H. Wills Physics Laboratory and Department of Electrical and Electronic Engineering, University of Bristol, Merchant Venturers Building, Woodland Road, Bristol BS8 1UB, UK}

\author{Mile Gu}
\affiliation{Center for Quantum Information, Institute for Interdisciplinary Information Sciences, Tsinghua University, Beijing, China}
\affiliation{Centre for Quantum Technologies, National University of Singapore, 3 Science Drive 2, 117543 Singapore}
\author{Kavan Modi}
\affiliation{School of Physics, Monash University, Victoria 3800, Australia}

\title{Power of One Bit of Quantum Information in Quantum Metrology}
\date{\today}

\begin{abstract} We construct a model of quantum metrology inspired by the computational model known as deterministic quantum computation with one quantum bit (DQC1). Using only one pure qubit together with $l$ fully-mixed qubits we obtain measurement precision at the standard quantum limit, which is typically obtained using the same number of uncorrelated qubits in fully-pure states. The standard quantum limit can be exceeded using an additional qubit, which adds only a small amount of purity.  We show that the discord in the final state vanishes only in the limit of attaining infinite precision for the parameter being estimated.
\end{abstract}

\maketitle

Quantum information science has transformed how we view many computation, communication, and precision-measurement tasks. Emerging quantum technologies promise to solve problems that are intractable or impossible using classical counterparts. However, in many cases the origins of quantum enhancements remain the subject of debate. Entanglement unambiguously plays a critical role in many tasks that use pure states, but this often ceases to be true when noise is added to the picture~\cite{arXiv:0201143}.  One of the most studied tasks that uses noisy qubits is provided by a model called DQC1, introduced by Knill and Laflamme~\cite{KL}.  DQC1 performs a specific type of classically-hard computation using highly-mixed quantum states, and thereby seriously challenges the notion that pure-state entanglement plays an essential role in quantum computation.

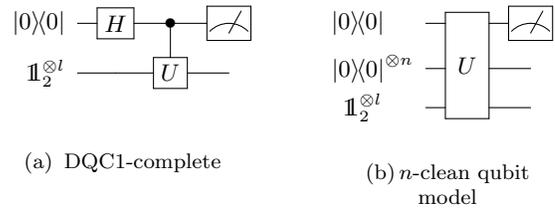
\begin{figure}
\subfigure[$\,$ DQC1-complete]{
\hspace{1cm}\Qcircuit @C=0.8em @R=0.7em {
\lstick{\ket{0}\!\!\bra{0}}  & \gate{H} & \ctrl{1} & \meter \\
\lstick{\openone_2^{\otimes l}}  & \qw & \gate{U} & \qw \\\\\\\\ }}
\hspace{1cm}
\subfigure[$\,n$-clean qubit model]{
\hspace{1cm}\Qcircuit @C=0.8em @R=0.7em {
\lstick{\ket{0}\!\!\bra{0}^{\quad \,}}  & \multigate{2}{U} & \meter \\
\lstick{\ket{0}\!\!\bra{0}^{\otimes n}} & \ghost{U} & \qw \\
\lstick{\openone_2^{\otimes l \quad \, \,}}  & \ghost{U} & \qw \\\\\\ }}
\caption{\emph{Circuits for DQC1}. (a) A DQC1-complete problem is to compute the normalised trace of a unitary transformation. For the circuit, a Hadamard gate is applied to the control qubit followed by controlled-unitary transformation on the register, and measurement of the control.  After several runs of the circuit, an estimate is obtained for $\braket{\sigma_x}+ i \braket{\sigma_y} = \tr (U)/{2^l}$. The precision of this estimate depends only on the number of runs of circuit. The protocol also works when the control qubit is partially pure at the start -- as given by the state given in Eq.~\eqref{epstate}.  In this case, the number of runs must be increased by a factor $1/\epsilon^2$ to achieve the same precision as when the control qubit is initially pure. (b) In the general DQC1 problem, an additional $n \sim \log(l)$ pure qubits can be introduced without altering the computational power of this model~\cite{Ambainis, shorjordan}. \label{dqc1}}\vspace{-10pt}
\end{figure}

The task performed by DQC1 is to estimate the normalised trace of a quantum circuit $U$ that acts on a collection of $l$ register qubits, as depicted in Fig.~\ref{dqc1}(a). The initial state comprises one ``clean'' pure qubit together with register qubits that are maximally mixed, and only unitary gates are used for the computation. Remarkably, the precision of the estimate does not scale with the size of $U$. It is intuitively clear that DQC1 achieves an exponential speedup over any classical algorithm which finds and sums the $2^l$ eigenvalues for $U$, and there is now considerable evidence which supports the existence of a true quantum speedup for DQC1~\cite{dattavidal}.  Several works have also analysed how the computational power of DQC1 changes as resources, such as additional pure qubits and measurements, are added~\cite{Ambainis, shorjordan, PhysRevLett.112.130502}, see Fig.~\ref{dqc1}(b).

Some studies have also investigated the role of entanglement and quantum discord~\cite{zurek,ollivierzurek} in the speedup achieved by DQC1~\cite{dattaflammiacaves, dattashaji}. It has been found that the discord generated at the output of DQC1 \footnote{Discord is asymmetric with respect to the subsystems for which it is defined. Here we refer to the discord for measurements on the control qubit, while the discord for measurements on the register is always zero when the register is fully mixed. The entanglement between these two partitions is zero when the register is fully mixed.}, by unitary transformations which are randomly selected according to the Haar measure, remains a fixed proportion of the maximum possible as the unitary transformations increase in size. However, the amount of entanglement generated by these unitary transformations is vanishing.  It is not yet known that happens to entanglement or discord at intermediate steps in a DQC1 computation. In contrast, it well known for pure-state quantum computation that unbounded entanglement is necessary for exponential speedups when using circuits composed of gates of bounded size~\cite{arXiv:0201143}.

We now turn to quantum metrology, and the problem of achieving quantum advantage for precision in the task of phase estimation, which is used for highly-sensitive measurements of physical parameters~\cite{dowling, natrev, photonictechnologies}. Phase-estimation strategies that cannot exploit quantum features are subject to the \emph{standard quantum limit} (SQL) for precision, given by $\Delta \phi = 1/\sqrt{n}$ where $n$ particles are used as probe and $\phi$ is the parameter to be estimated. For example, this limit applies when $n$ single pure qubits in the $\ket{+}= (\ket{0}+\ket{1})/\sqrt{2}$ state are used to measure the phase for a Pauli rotation
\begin{gather}\label{unitary}
u_\phi = e^{i\phi \mathfrak{g}} \quad \mbox{where} \quad \mathfrak{g}=\ket{1}\!\!\bra{1}.
\end{gather}
However, when a GHZ state $\ket{+_n} = (\ket{0}^{\otimes n} +\ket{1}^{\otimes n})/\sqrt{2}$ is used as the probe state with $G=\sum_{j=1}^n \mathfrak{g}_j$, the precision scales at the Heisenberg limit $\Delta \phi = 1/n$, which is the best precision achievable~\cite{natrev}.

Inspired by DQC1, we ask whether a large ensemble of mixed qubits can be used as the basis of a powerful sensor. We consider a model where only one (or few) clean qubits are accessible, and only one qubit can be measured at the end \footnote{Inclusion of additional measurements would allow for the preparation of pure states  from mixed qubits, which fundamentally change the power of the model.}.  Physical systems where our model is most relevant include NMR~\cite{OxfordSci, simmonsPRA, SchaffryPRA} and some cold-atom systems~\cite{EITRydbergGate, mansellbergamini}. For these systems often only bulk operations on the register qubits are available --- which is so say the same operation is applied to every register qubit, optionally under global control. Hence, we add a bulk-operations constraint to our model.

For mixed-state models of phase estimation, recent results challenge any presupposed link between entanglement and quantum advantage for measurement precision.  Ref.~\cite{PhysRevA.77.052320} considers an algorithm for multi-parameter estimation using DQC1.  This algorithm uses an adaptive protocol based on a series of estimates with different interactions times, to achieve a final precision scaling with the inverse total interaction time. Ref.~\cite{prx} analyses the situation where a unitary circuit is used to prepare probe states from $n$ uncorrelated qubits in the state $\rho_{\epsilon}$ given below in Eq.~\eqref{epstate}.  Strategies using probe states with only classical correlations~\cite{henderson01a}, which is to say they are diagonal in the $\sigma_z$ basis up to local-unitary transformations~\cite{arXiv:1005.4348}, are compared with strategies with exploit entanglement and quantum discord (defined as in~\cite{modietal}).  It was found that circuits which generate non-classical correlations can achieve a quadratic quantum advantage compared to circuits generating only classical correlations at fixed $\epsilon$ in Eq. \eqref{epstate}. This result holds even for small values of $\epsilon$ where there is no entanglement but large amounts of discord, and the amount of discord also grows with $n$. Another recent analysis considers phase estimation using an interferometer, where the spectrum of the interferometer Hamiltonian is fixed but not its eigenbasis. The authors found that the minimum amount of statistical information that can be extracted about the unknown phase in the problem also constitutes a measure of the discord-type correlations in the probe state~\cite{adesso}.


\emph{Parameter estimation.}--- In the classical theory for parameter estimation~\cite{kullback} a probability distribution $\mathbf{p}$ is subjected to a process that is a function of a single parameter $\phi$. The process alters the initial distribution $\mathbf{p}$ into $\mathbf{p}(\phi)$, which depends on the value of $\phi$. Differentiating between the initial and the final distributions allows for the determination of the value of $\phi$. The uncertainty in this value is bounded by the Fisher information, which is given by:
\begin{gather}\label{clF}
\Delta \phi \ge \frac{1}{\sqrt{F}}, \quad \mbox{with} \quad
F = \sum_k \frac{[\partial_\phi p_k(\phi)]^2}{p_k(\phi)}
\end{gather}
and $p_k$ is the probability for observing outcome $k$. The above inequality is the Cram\'{e}r-Rao bound~\cite{cramer, rao}.

When using a quantum system, the initial and final probability distributions are replaced by density operators $\sigma$ and $\sigma(\phi)$ respectively. The final state is measured by a positive operator valued measure (POVM) $\{\Pi_k\}$ to yield classical probabilities $p_k = \tr[\Pi_k \sigma(\phi)]$, from which $F$ in Eq.~\eqref{clF} can be computed.  If the process which is parameterized by $\phi$ is unitary, then the Fisher information when optimised over all POVMs is given by the quantum Fisher information~\cite{braunsteincaves}:
\begin{gather}\label{qF}
F_{\mbox{q}}= 4 \sum_{i>j} \frac{(\lambda_i - \lambda_j)^2}{\lambda_i + \lambda_j} \left\vert \braket{\psi_i \left| G \right| \psi_j} \right\vert^2,
\end{gather}
where $\{\lambda_i\}$ are the eigenvalues,$\{\ket{\psi_i}\}$ are the eigenvectors of $\sigma$, and $G$ is the Hamiltonian generator of the phase shift.  This formula for $F_{\mbox{q}}$ is very powerful: It yields the lower bound for the precision of $\phi$ without needing the explicit form of the optimising POVM, and enables a straightforward comparison between different initial states.

\begin{figure}
	\setlength{\unitlength}{5cm}
	\begin{picture}(0,0)
	\put(0.248,-.39){\line(0,1){.045}} %
	\put(0.248,-.56){\line(0,1){.045}} %
\end{picture}
\begin{center}
\hspace{1cm}\Qcircuit @C=0.8em @R=0.7em {
\lstick{\ket{0}\!\!\bra{0}^{\,\,\,\,\,\,\,\,}} & \qw  & \gate{H} & \ctrl{3} & \gate{u_\phi}  & \ctrl{3} & \ctrl{1} & \gate{v_{\theta(r)}} & \meter \\
\lstick{\ket{0}\!\!\bra{0}^{\otimes n}} & \qw & \qw & \targ & \gate{u^{\otimes n}_\phi} & \targ & \gate{v^{\otimes n}_{\theta(r)}} & \qw & \\
\lstick{\rho_\epsilon^{\otimes m\,\,\,\,\,\,\,}} & \qw & \qw & \targ    & \gate{u^{\otimes m}_\phi} & \targ & \gate{v^{\otimes m}_{\theta(r)} } & \qw & \\
\lstick{\openone_2^{\otimes l\,\,\,\,\,\,\,\,}} & \qw & \qw & \targ    & \gate{u^{\otimes l}_\phi} & \targ & \gate{v^{\otimes l}_{\theta(r)} } &  \qw & \\}
\end{center}
\caption{Illustration of our general scheme for phase estimation:  There are $n$ pure qubits, $m$ semi-pure qubits, and $l$ fully-mixed qubits ($\openone_2 = \openone /2$).  The first qubit is the control, and the remaining qubits constitute the register.  Only bulk operations are permitted for gates used to prepare the probe state and implement the readout procedure.  A bulk \textsc{Cnot} is used to prepare the probe, and then each qubit is subjected to the unitary operation given in Eq.~\eqref{unitary} for which $\phi$ is to be determined. The readout procedure is adaptive: $\theta(r)$ is the estimate for $\phi$ after the first $r\!-\!1$ rounds. This value is used to configure the readout circuit, which is a \textsc{Cnot} followed by controlled-$v_{\theta(r)}$ on all qubits of the register, and measurement on the control. 
 \label{fig:metrology}}
\end{figure}
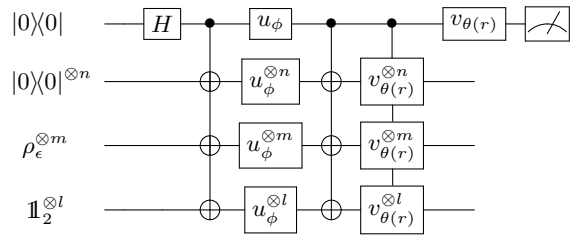


\emph{The setup.}--- Our model uses three registers: one with $n$ pure qubits; one with $m$ qubits with finite purity as given in Eq. \eqref{epstate}; and one with $l$ fully-mixed qubits. Along with these three registers, there is one pure qubit in state $\ket{0}$ which serves as the control.  The total initial state is
\begin{gather}\label{statezero}
\sigma_0 = \ket{0}\!\!\bra{0} \otimes \ket{0}\!\!\bra{0}^{\otimes n} \otimes \rho_\epsilon^{\otimes m} \otimes \frac{\openone^{\otimes l}}{2^l}.
\end{gather}
with
\begin{gather}\label{epstate}
\rho_\epsilon = \frac{1}{2}\begin{pmatrix} 1+ \epsilon & 0 \cr 0 & 1-\epsilon \cr \end{pmatrix}, \quad 0 < \epsilon < 1.
\end{gather}
(Later we will take the limit of $n,m \to 0$ to develop compare our results with DQC1.) To prepare the probe state $\sigma$ we apply the Hadamard gate to the control qubit followed by a \textsc{Cnot} gate for all qubits in the register.  Next each qubit in the register is allowed to evolve freely under the unitary operation given in Eq.~\eqref{unitary}.  The readout procedure consists of another controlled operation and measurement of the control qubit. The full protocol is shown in Fig.~\ref{fig:metrology}.

To compute $F_{\mbox{q}}$ for $\sigma$ above, we note that $\sigma_0$ has eigenvectors of the form $\ket{\pm ; B^{n}_0; B^{m}_j ; B^{l}_k}$; here $B^{a}_i$ represents a binary string of length $a$ with 1s appearing $i$ times, and the semicolons separate the control qubit and the three registers. There are $\binom{m}{j}\binom{l}{k}$ such eigenvectors each with eigenvalue
\begin{gather}
\lambda_{j}^{+} = \frac{1}{2^{m+l}}(1+ \epsilon)^{m-j}(1-\epsilon)^j
\end{gather}
when the control qubit is in state $\ket{+}$, and the eigenvalue is $\lambda_j^{-}=0$ otherwise. After the first \textsc{Cnot} gate the eigenvectors are
\begin{gather}
\ket{\psi_{jk}^{\pm}} = \frac{1}{\sqrt{2}} \left(\ket{0; B^{n}_0; B^{m}_j ; B^{l}_k} \pm \ket{1 ; C^{n}_0; C^{m}_j ; C^{l}_k} \right),
\end{gather}
where $C^{a}_j$ is the \textsc{not} of $B^{a}_j$, i.e., $\ket{C^{a}_j} = \sigma_x^{\otimes a}\ket{B^{a}_j}$.

The generator of the phase shift is $G = \sum_x \ket{1}\!\!\bra{1}_x \otimes \openone_{\bar{x}},$ where $\openone_{\bar{x}}$ is the identity operator on all but $x$th qubit, and $x$ runs from 1 to $n+m+l$. Next, we note that the components of the eigenstate of the prepared state are eigenvectors $G$:
\begin{align}
&G \ket{0;B^n_0;B^m_j;B^l_k} = (j+k) \ket{0;B^n_0;B^m_j;B^l_k}, \\
&G \ket{1; C^n_0; C^m_j; C^l_k} =  \notag
 (n + 1 + m - j + l - k) \\& \hspace{3.5cm}
\times \ket{1; C^n_0; C^m_j; C^l_k}
\end{align}
and therefore
\begin{align}
\braket{\psi_{j'k'}^\pm \vert G  \vert \psi_{jk}^+}
=& \frac{1}{2}(j+k) \braket{B^{'m}_{j'} \vert B^m_j} \braket{B^{'l}_{k'} \vert B^l_k} \notag\\
&\pm \frac{1}{2} {(n+m+l-j-k+1)} \notag\\
& \quad \times \braket{C^{'m}_{j'} \vert C^m_j} \braket{C^{'l}_{k'} \vert C^l_k} .
\label{eqinnerprod}
\end{align}
Eq.~\eqref{eqinnerprod} is non-zero only when $j=j'$ and $k=k'$.  We note that the numerator of the first term in Eq.~\eqref{qF} is the difference in two eigenvalues, and therefore it is only necessary to consider $\braket{\psi_{j,k}^- \vert G  \vert \psi_{j,k}^+}$.  Hence,
\begin{align}
F_{\mbox{q}} =&\,\, 4 \sum_{j=0}^{m} \binom{m}{j} \lambda_j^+
\sum_{k=0}^{l} \binom{l}{k} 
  \; \left \vert  \braket{\psi_{j,k}^- \vert G  \vert \psi_{j,k}^+} \right\vert^2 \notag \\
\label{1qm.2} =&\,\, l+m(1-\epsilon^2)+(1+n+\epsilon m)^2\,.
\end{align}

We can make several observations concerning Eq.~\eqref{1qm.2}: (i) $F_{\mbox{q}}$ is always greater or equal to the SQL value, which is $1+l+m+n$. (ii) The SQL is attained when $m=n=0$, i.e. the case which is analogous to DQC1 \footnote{If the control qubit is assumed to have initial state given by Eq.~\eqref{epstate} then $F_{\mbox{q}}$ is $\epsilon^2 l$, i.e. just as in DQC1 there is an overhead scaling with $\epsilon^2$.}. (iii) If $\epsilon$ is small (or even 0) there is a linear contribution of $m$ corresponding to size of the register of partially-pure qubits. (iv) $(n+1)^2$ exhibits the well-known quadratic enhancement for entangled pure states of $n+1$ qubits, and there is an additional contribution equivalent to $\epsilon m$ extra pure qubits.


\begin{figure}
\includegraphics[width=0.4\textwidth]{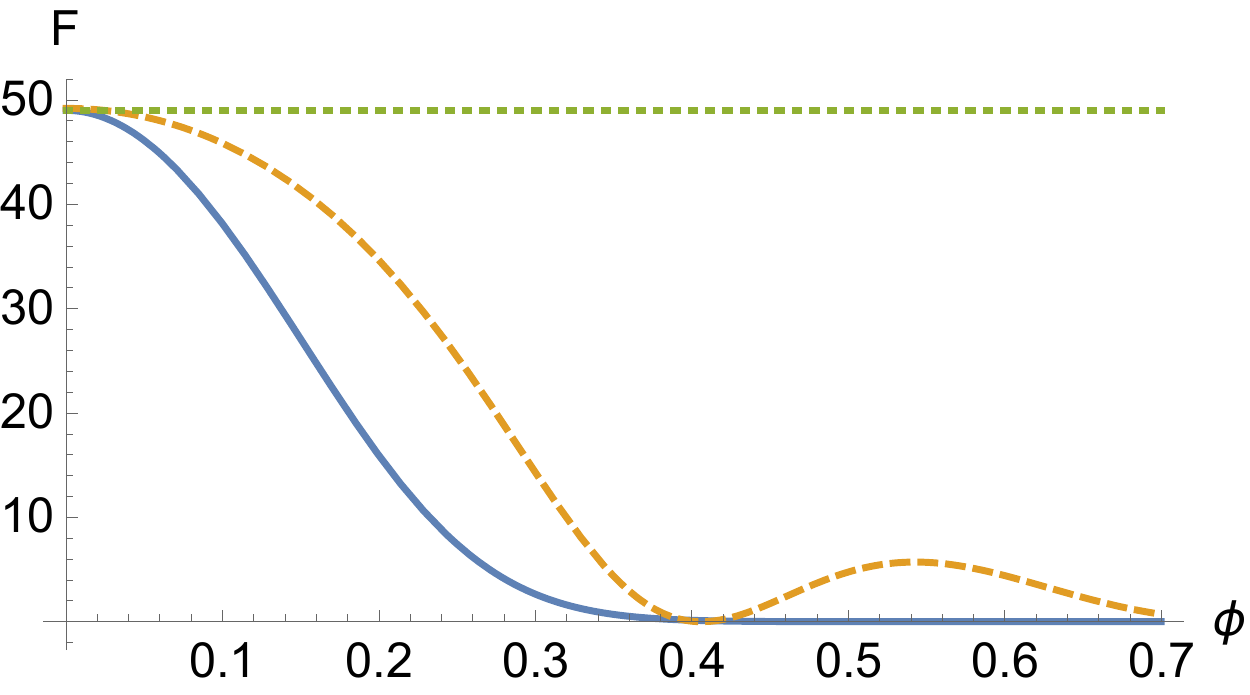}
\caption{(Colour online.) Fisher information, $F$, computed from the probability distribution in Eq.~\eqref{probdist} for $n=6$ and $m=l=0$ (dotted); $n=l=0$ and $m=11$ with $\epsilon= 0.49$ (dashed); and $l=48$ and $n=m=0$ (solid). $F$ is independent of $\phi$ when register has no mixed qubits. All three cases have the same value for the quantum Fisher information $F_{\mbox{q}}$. \label{fig:fisher}}
\end{figure}

\emph{Readout procedure.---} Apart from preparation of the probe state, attention must be given to the bulk-operation requirements for implementing the measurements for the readout procedure. In other words, we need to consider how $F_{\mbox{q}}$ given in Eq.~\eqref{1qm.2} can be attained via a suitable POVM, which in general can require entangled measurements~\cite{njp}. For our model, the following method suffices (illustrated in Fig.~\ref{fig:metrology}): a bulk \textsc{Cnot} gate is performed, followed by a bulk controlled-$v_{\theta(r)}$ where $v_{\theta(r)} = \exp\left\{-i  \theta(r) \sigma_z \right\} $, and a measurement on the control qubit. $\theta(r)$ here is taken to be the estimate of $\phi$ following the $(r-1)$th round.  The initial estimate $\theta(0)$ can assume no prior knowledge of $\phi$. In each successive round our estimate for $\phi$ is improved, i.e., $ \vert \theta(r)\!-\!\phi \vert < \vert \theta(r-1)\!-\!\phi \vert$, using an adaptive Bayesian update or maximum-likelihood method to maximize sensitivity~\cite{Xiang}.

The measurement of the control qubit along the $\sigma_x$ direction yields probability distribution
\begin{align}\label{probdist}
q^\pm(r) =&\, \frac{1}{2}\left( 1 \pm x(r) \right) \quad \mbox{where} \\ \notag
x(r) =&\, \mbox{Re}\Big\{ e^{i (n+1) \omega(r)} \cos^l(\omega(r)) \\
& \times \left[ \cos(\omega(r)) + i \epsilon \sin(\omega(r))\right]^m \Big\},
\end{align}
and $\omega(r) =\theta(r) - \phi$. The value for $F$ computed from this probability distribution, using Eq.~\eqref{clF}, yields a value that approaches the quantum Fisher information in Eq.~\eqref{1qm.2} as $\omega(r) \approx 0$. That means that the adaptive protocol described above will yield the optimal Fisher information as the estimate $\theta(r)$ approaches $\phi$. We have plotted three cases in Fig.~\ref{fig:fisher}. 


\emph{One+$\epsilon$ clean qubit metrology.---} It is clear that the metrology protocol presented in Fig.~\ref{fig:metrology} is a special case of DQC1 computation, provided $n+m \sim \log(l)$. In fact, we can think of the circuit as an application of a bulk controlled-unitary operation $^cU_{\omega(r)}$, where the unitary operation is
\begin{gather}\label{metuni}
U_{\omega(r)} = \left(u_\phi^\dag v_{\theta(r)} \sigma_x \, u_\phi \, \sigma_x \right)^{\otimes l+m+n}.
\end{gather}
Using this feature we will examine the discord hypothesis for DQC1 due to~\cite{dattashaji}. 

However, let us first consider the case where $n=0$, $m=1$, and $l > 0 $. In this case the probe state $\sigma$ is entangled for any value of $\epsilon > 0$. This can be understood by noting that a non-positive partial transpose for $\sigma$ of the state results from applying a \textsc{Cnot} gate on the state in Eq.~\eqref{epstate}, controlled on $\ket{+}$. Another way to see this is by noting that the value for $F_{\mbox{q}}$ here beats the SQL~\cite{lloydPRL}:
\begin{gather}
F_{\mbox{q}} = l+ 2 + 2 \epsilon > l + 2.
\end{gather}
In other words, even with one qubit with finite purity we can attain better precision than what is possible classically. Adding more qubits to the registers for initially partially-mixed and pure qubits, the entanglement (between the control and registers) will increase as well as the value for $F_{\mbox{q}}$.

\emph{One-pure-qubit metrology.---} We now let $n=m=0$, i.e., consider a $l+1$ qubit state with only one pure qubit and $l$ qubits in fully-mixed state. From Eq.~\eqref{1qm.2} we see that  $F_{\mbox{q}}$ has the SQL value of $l+1$ qubits, and that the SQL is attained using only one pure qubit and $l$ fully-mixed qubits. This is highly counterintuitive in the classical setting, where completely-mixed states cannot be used to yield additional information from a phase measurement,
and the maximum value for $F$ would be $1$ (as attained by a single pure qubit).  Therefore the enhancement of $F_{\mbox{q}}$ by $l$ is an entirely quantum-mechanical phenomenon.

It is tempting to say that the resource that enables this enhancement in $F_{\mbox{q}}$ is the entanglement or discord in $\sigma$. However, a closer look at $\sigma$ in the limit $\epsilon \to 0$ reveals that it is an equal mixture of products of eigenstates of $\sigma_x$ (for which $\ket{-}$ occurs even number of times),
\begin{gather}
\sigma = \frac{1}{2^{l+1} }\left(\openone^{\otimes l+1} + \sigma_x^{\otimes l+1} \right),
\end{gather}
and it is therefore fully-classically correlated  
~\cite{arXiv:1005.4348}.
Though $\sigma$ is separable, and therefore preparable via unrestricted LOCC, it cannot be prepared using bulk LOCC operation. Without the \textsc{Cnot} gate used in the state preparation, which is controlled on a quantum superposition, the register of maximally-mixed qubits cannot be exploited.

At this point we can ask whether there is any discord present in the final state of the circuit.  In Ref.~\cite{arXiv:1004.0190} it was shown that there is no discord in the output state of a DQC1 circuit when the controlled-unitary operation is Hermitian, i.e. $U = U^\dag$ in Fig.~\ref{dqc1} (see also Refs.~\cite{rmp} and for further details). The unitary operator $U_{\omega(r)}$, in Eq.~\eqref{metuni}, is Hermitian if and only if $\omega(r) = 0$, i.e. when $\phi$ is known to perfect precision.  Therefore it may be observed that the circuit in Fig.~\ref{fig:metrology} contains discord for all runs except when $\phi$ is fully known. Repeating this analysis for arbitrary values of $l,\,m,\,n>0$ shows that the final state is always separable, but has finite discord except when $\omega(r) = 0$. The only exception is when $l=m=0$, in which case the final state has no correlations. We may conclude that noisy input states lead to discordant output states in our model, which sheds new light on the constant level of discord at the output of DQC1 found in Ref.~\cite{dattashaji}.

\emph{Discussion.---} We have constructed a model of quantum metrology, inspired by DQC1, that uses highly-mixed states as its enabling resource.  Our most surprising result arises when the register is taken to be fully mixed.  In this case, the probe state is classical correlated, and yet it can only be prepared via a coherent quantum interaction due to the bulk operation constraint of our model.  Whilst there is no entanglement or discord in the probe state, we have found that the state at the output always has discord, except in the limit of infinite precision for the phase parameter being estimated.  Our model then surpasses the performance of a classical setup when only one qubit in the register has a finite amount of purity.  In this case the probe state also has entanglement, which is widely understood to be essential for achieving precision beyond the SQL. 

Our results provide support for both entanglement and discord as enabling quantum resources in quantum metrology. Perhaps more importantly, our model shows how a large ensemble of highly-mixed quantum systems can be of great utility for quantum sensing. Since our model only requires bulk coherent operations on the ensemble, it has the potential to enable a scalable quantum technology could challenge state-of-the-art classical sensors in the near future. The biggest practical weakness of our model lies in the fact that if even a single qubit is lost between the first and last controlled gates, all sensitivity is lost --- a problem which is shared by any measurement device using pure GHZ states or NOON states in the context of interferometry~\cite{dowling}.

{\bf Acknowledgements.} H.C. is grateful for financial support from the University of Bristol. M.G is supported by the John Templeton Foundation 54914, the National Basic Research Program of China Grant 2011CBA00300, 2011CBA00302 and the National Natural Science Foundation of China Grant 11450110058, 61033001, 61361136003. 

\bibliography{metrology}

\newpage
\appendix
\begin{widetext}
\section{State evolution throughout the protocol}
\noindent We begin with the initial state
\begin{align}
\sigma_0 = \frac{1}{2} 
\left(\begin{matrix}
\ket{0}\!\!\bra{0}^{\otimes n} \otimes \rho_\epsilon^{\otimes m} \otimes \openone_2^{\otimes l} & \ket{0}\!\!\bra{0}^{\otimes n} \otimes \rho_\epsilon^{\otimes m} \otimes \openone_2^{\otimes l}\\
\ket{0}\!\!\bra{0}^{\otimes n} \otimes \rho_\epsilon^{\otimes m} \otimes \openone_2^{\otimes l}&
\ket{0}\!\!\bra{0}^{\otimes n} \otimes \rho_\epsilon^{\otimes m} \otimes \openone_2^{\otimes l}
\end{matrix}\right).
\end{align}
After the first \textsc{Cnot} we have
\begin{align}
\sigma_1 = \frac{1}{2^{l+1}} 
\left(\begin{matrix}
\ket{0}\!\!\bra{0}^{\otimes n} \otimes \rho_\epsilon^{\otimes m} \otimes \openone^{\otimes l} &\ket{0}\!\!\bra{1}^{\otimes n} \otimes (\rho_\epsilon\sigma_x)^{\otimes m} \otimes \sigma_x^{\otimes l} \\
\ket{1}\!\!\bra{0}^{\otimes n} \otimes (\sigma_x\rho_\epsilon)^{\otimes m} \otimes \sigma_x^{\otimes l}
& \ket{1}\!\!\bra{1}^{\otimes n} \otimes (\sigma_x\rho_\epsilon\sigma_x)^{\otimes m} \otimes \openone^{\otimes l}
\end{matrix}\right).
\end{align}
Next the phase is encoded, but note that $u_\phi$ commutes with $\ket{0}\!\!\bra{0}$, $\rho_\epsilon$, and $
\sigma_x \rho_\epsilon \sigma_x$
\begin{align}
\sigma_2 = \frac{1}{2^{l+1}} 
\left(\begin{matrix}
\ket{0}\!\!\bra{0}^{\otimes n} \otimes \rho_\epsilon^{\otimes m} \otimes \openone^{\otimes l} &
e^{-i (n+1) \phi}
\ket{0}\!\!\bra{1}^{\otimes n} \otimes (u_\phi \rho_\epsilon \sigma_x u_\phi^\dag )^{\otimes m} \otimes (u_\phi \sigma_x u_\phi^\dag)^{\otimes l}\\
e^{i(n+1)\phi}
\ket{1}\!\!\bra{0}^{\otimes n} \otimes (u_\phi\sigma_x \rho_\epsilon u_\phi^\dag)^{\otimes m} \otimes (u_\phi\sigma_x u_\phi^\dag)^{\otimes l}& 
\ket{1}\!\!\bra{1}^{\otimes n} \otimes (\sigma_x\rho_\epsilon\sigma_x)^{\otimes m} \otimes \openone^{\otimes l}
\end{matrix}\right).\notag
\end{align}
Next we apply the second \textsc{Cnot} gate 
\begin{align}
\sigma_3 = \frac{1}{2^{l+1}} 
\left(\begin{matrix}
 \rho_\epsilon^{\otimes m} \otimes \openone^{\otimes l} & e^{-i (n+1) \phi}
 (u_\phi \rho_\epsilon \sigma_x u_\phi^\dag \sigma_x)^{\otimes m} \otimes (u_\phi \sigma_x u_\phi^\dag \sigma_x)^{\otimes l} \\
e^{i(n+1)\phi} (\sigma_x u_\phi\sigma_x \rho_\epsilon u_\phi^\dag )^{\otimes m} \otimes (\sigma_x u_\phi\sigma_x u_\phi^\dag)^{\otimes l}
& \rho_\epsilon^{\otimes m} \otimes \openone^{\otimes l} \end{matrix}\right) \otimes \ket{0}\!\!\bra{0}^{\otimes n} . \notag
\end{align}
Finally we apply controlled-$v_{\theta(r)}$ as well as $v_{\theta(r)}$ on the control qubit. Note that $v_{\theta(r)}$ also commutes with $\ket{0}\!\!\bra{0}$, $\rho_\epsilon$, and $\sigma_x \rho_\epsilon \sigma_x$. Measuring the control qubit in the basis of $\sigma_x$ gives us
\begin{align}
q^{\pm}(r) = \braket{\pm \vert \sigma_4 \vert \pm} =& 
\frac{1}{2}\left(1 \pm \mbox{Re} \{\tr[e^{-i (n+1) (\phi-\theta(r))}
\ket{0}\!\!\bra{0}^{\otimes n} \otimes (\rho_\epsilon u_\phi \sigma_x u_\phi^\dag \sigma_x v_{\theta(r)}^\dag)^{\otimes m} \otimes (u_\phi \sigma_x u_\phi^\dag \sigma_x v_{\theta(r)}^\dag)^{\otimes l}/2^l] \} \right)\\
=& \frac{1}{2} \left( 1 \pm \mbox{Re} \{e^{-i (n+1) (\phi-\theta(r))}
\tr[\rho_\epsilon u_\phi \sigma_x u_\phi^\dag \sigma_x v_{\theta(r)}^\dag]^{m} \tr[u_\phi \sigma_x u_\phi^\dag \sigma_x v_{\theta(r)}^\dag]^{l}/2^l \}\right)\\
=& \frac{1}{2} \left(1 \pm \mbox{Re}\left\{ e^{i (n+1) \omega(r)} \left[ \cos(\omega(r)) + i \epsilon \sin(\omega(r))\right]^m \cos^l(\omega(r)) \right\}\right).
\end{align}
\end{widetext}
\end{document}